\documentclass[letterpaper, 10 pt, conference]{IEEEtran}
\IEEEoverridecommandlockouts
\DeclareUnicodeCharacter{0301}{\'{i}}
\usepackage{comment}
\usepackage{caption}
\usepackage{subcaption}
\usepackage[font=small]{caption}
\usepackage{multirow, makecell}
\usepackage{amsmath,amssymb,amsfonts}
\usepackage{algorithmic}
\usepackage{wasysym}
\usepackage{graphicx}
\usepackage{textcomp}
\usepackage{xcolor}
\usepackage{dblfloatfix}
\makeatletter
\usepackage{hyperref}
\newcommand\notsotiny{\@setfontsize\notsotiny{6.83415}{7.6928}}
\makeatother
\def\BibTeX{{\rm B\kern-.05em{\sc i\kern-.025em b}\kern-.08em
    T\kern-.1667em\lower.7ex\hbox{E}\kern-.125emX}}

\makeatletter 
\newcommand{\linebreakand}{%
  \end{@IEEEauthorhalign}
  \hfill\mbox{}\par
  \mbox{}\hfill\begin{@IEEEauthorhalign}
}
\makeatother 

\usepackage[
style=ieee
]{biblatex}
\AtBeginBibliography{\notsotiny}
\begin{document}

\title{Assessing gender fairness in EEG-based machine learning detection of Parkinson's disease: \\ A multi-center study \\
}

\author{%
  \IEEEauthorblockN{%
    \parbox{\linewidth}{\centering
      Anna Kurbatskaya\IEEEauthorrefmark{1},
      Alberto Jaramillo-Jimenez\IEEEauthorrefmark{2}\IEEEauthorrefmark{3}\IEEEauthorrefmark{4},
      John Fredy Ochoa-Gomez\IEEEauthorrefmark{5}, \\
      Kolbjørn Brønnick\IEEEauthorrefmark{2}\IEEEauthorrefmark{3} and
      Alvaro Fernandez-Quilez\IEEEauthorrefmark{1}\IEEEauthorrefmark{6}%
    }%
  }%
  \\
  \IEEEauthorblockA{%
    \IEEEauthorrefmark{1}\footnotesize Department of Electrical Engineering and Computer Science, University of Stavanger, Stavanger, Norway\\
    \IEEEauthorrefmark{2}Centre for Age-Related Medicine (SESAM), Stavanger University Hospital, Stavanger, Norway \\
    \IEEEauthorrefmark{3}Department of Public Health, Faculty of Health Sciences, University of Stavanger, Stavanger, Norway \\
    \IEEEauthorrefmark{4}Grupo de Neurociencias de Antioquia, Universidad de Antioquia, School of Medicine, Medellín, Colombia \\
    \IEEEauthorrefmark{5}Grupo Neuropsicolog\'{i}a y Conducta, Universidad de Antioquia, School of Medicine, Medellín, Colombia \\
    \IEEEauthorrefmark{6}Stavanger Medical Imaging Laboratory (SMIL), Department of Radiology, Stavanger University Hospital, Stavanger, Norway \\
    Email: \IEEEauthorrefmark{1}anna.kurbatskaya@uis.no,
    \IEEEauthorrefmark{2}\IEEEauthorrefmark{3}\IEEEauthorrefmark{4}alberto.jaramilloj@udea.edu.co,
    \IEEEauthorrefmark{5}john.ochoa@udea.edu.co,
    \IEEEauthorrefmark{2}\IEEEauthorrefmark{3}kolbjorn.s.bronnick@uis.no,
    \IEEEauthorrefmark{1}\IEEEauthorrefmark{6}alvaro.f.quilez@uis.no%
  }%
}

\maketitle

\begin{abstract}

As the number of automatic tools based on machine learning (ML) and resting-state electroencephalography (rs-EEG) for Parkinson's disease (PD) detection keeps growing, the assessment of possible exacerbation of health disparities by means of fairness and bias analysis becomes more relevant. Protected attributes, such as gender, play an important role in PD diagnosis development. However, analysis of sub-group populations stemming from different genders is seldom taken into consideration in ML models' development or the performance assessment for PD detection. In this work, we perform a systematic analysis of the detection ability for gender sub-groups in a multi-center setting of a previously developed ML algorithm based on power spectral density (PSD) features of rs-EEG. We find significant differences in the PD detection ability for males and females at testing time (80.5$\%$ vs. 63.7$\%$ accuracy) and significantly higher activity for a set of parietal and frontal EEG channels and frequency sub-bands for PD and non-PD males that might explain the differences in the PD detection ability for the gender sub-groups.

\end{abstract}

\begin{IEEEkeywords}
fairness analysis, Parkinson’s disease, EEG, machine learning, classification, multi-center
\end{IEEEkeywords}

\section{Introduction}
Machine learning (ML) advances are facilitating and accelerating the development of automatic tools aimed to aid clinicians in their daily tasks \cite{Davenport2019}. As the number of ML tools continues to grow and affect high-stake decisions, \textit{fairness analysis} and \textit{bias assessment} are becoming equally important to ensure that algorithms do not create or amplify existing biases in the population under analysis. On the contrary, calls have been made to emphasize getting an equally accurate performance for sub-populations stemming from, for example, different genders or ethnicities \cite{Char2018, Fernandez2022}. 

Parkinson's disease (PD) is a slowly progressive neurodegenerative disorder, primarily manifesting in typical parkinsonian motor symptoms \cite{Kalia2015}. Currently, no diagnostic tests allow clinicians to make a definitive early diagnosis. Electroencephalography (EEG) presents a non-invasive and low-cost alternative method that has been shown to be a reliable clinical research tool for PD \cite{Babiloni2020}. In particular, quantitative EEG (qEEG) and spectral features, such as power spectral density (PSD), have garnered considerable interest due to the association of their changes with the progression of PD \cite{Neufeld1994, Pezard2001}.

Gender, along with aging, genetics, environment, and the immune system, play an essential role in the development and phenotypical expression of PD \cite{Cerri2019}. Research has shown that the risk of PD is higher for men \cite{Baldereschi2000}, although there is a higher mortality rate and faster progression of the disease for women \cite{Dahodwala2018}. Further, men and women with PD tend to have different motor and non-motor profiles \cite{Miller2010}. For instance, imaging studies have reported significant differences for PD males and females in executive functions, more atrophy in cortical regions for males, cortical thinning in post-central and pre-central regions \cite{Oltra2022}, and reduced global cognition for PD males than females \cite{Reekes2020}.

Although a considerable amount of work has been carried out in rs-EEG and ML for PD \cite{Yuvaraj2018, Vanneste2018}, fairness and sub-group analysis of the developed tools is seldom conducted, despite the relevance of \textit{protected attributes} such as \textit{gender} \cite{Baldereschi2000}. Works in other fields, such as medical imaging, have already shown significant disparities in the diagnosis performance across \textit{datasets}, \textit{tasks}, and \textit{protected attributes} \cite{Seyyed2021}. For example, a significant underdiagnosis was found in chest X-ray classification algorithms for black females, potentially resulting in delays in treatment \cite{Seyyed2021}. Moreover, a significant decrease in diagnosis performance was reported for different genders using chest X-ray images \cite{Larrazabal2020}.

\begin{figure*}[!t]
    \centering
    \includegraphics[width=0.9\textwidth]{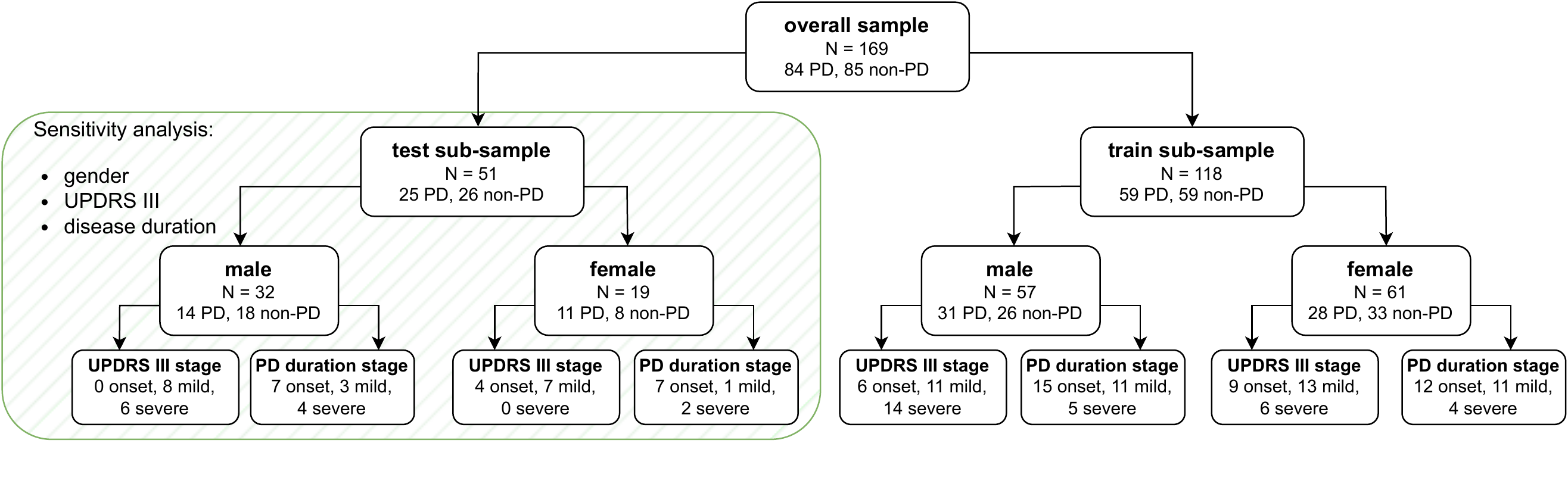}
    \vspace{-0.5cm}
    \caption{\small Breakdown of subject classes in train ($N$=118) and test ($N$=51) sub-samples. The population is divided into two gender sub-groups (males and females) followed by a division into disease-stage groups based on UPDRS III stage scores and disease duration in months.}
    \label{fig:distribution}
    \vspace{-0.6cm}
\end{figure*}

In this work, we systematically study the detection ability for the different genders of a previously developed PD detection model based on rs-EEG and PSD features \cite{Kurbatskaya2023}, where the analysis of gender sub-groups is motivated by their relevance from a clinical perspective. To the best of our knowledge, \textit{this is the first study attempting to understand the effect of a protected attribute such as gender in rs-EEG and ML models}. 

We make the following contributions: (i) a systematic analysis of an ML model performance \cite{Kurbatskaya2023} trained in a gender-balanced regime for different gender sub-populations (Fig. \ref{fig:distribution}); (ii) a qualitative analysis of the influence of Unified Parkinson's Disease Rating Scale (UPDRS) III score and PD duration on the exhibited disparities; (iii) an analysis of the contribution of the different EEG channels and frequency sub-bands in the disparities. We make our code publicly available on \href{https://github.com/Biomedical-Data-Analysis-Laboratory/multicentric-ML-Parkinson-detection}{GitHub}. 

We find significant differences in the detection ability for males and females at testing time. Further, we find that those differences might be explained by higher activity in a set of frequency sub-bands of CP1, CP2, CP5, CP6, Fp1, Fp2, and P3 channels for PD males when compared to PD females and in CP5, FC2, Fp1, and Fp2 for non-PD males and females. Our findings suggest that different frequency sub-bands and channels might have a higher expressiveness for PD detection depending on gender. 
\vspace{-0.1cm}

\section{Materials and Methods}
\vspace{-0.1cm}
\subsection{Data}
\subsubsection{Study population} We include rs-EEG recordings of 169 subjects (84 PD and 85 non-PD) from four datasets collected at different research centers in three countries: Colombia (Medell\'{i}n) \cite{Carmona_Arroyave2019}, Finland (Turku) \cite{Railo2020}, and the USA (Iowa City and San Diego) \cite{Anjum2020, Rockhill2021}. PD and non-PD subjects were matched by age in all the datasets. Furthermore, subjects were matched by gender, level of education, and cognitive performance in all the datasets except Turku.

\begin{table}[!ht]
\caption{\small Demographic and clinical characteristics (mean$\pm$SD) of PD and non-PD subjects of each dataset. y=years, m=months.}
\label{tab:datasets}
\begin{center}
\notsotiny
\resizebox{0.49\textwidth}{!}{\begin{tabular}{cccccc}
\hline
& Subjects & Age, y &  \# $\female$ (\%) & Duration, m & UPDRS III \\
    \hline
    \multirow{2}{*}{\textit{Iowa City}} & PD ($n$=14) & 70.5 (8.6) & 8 (57.2) & 66.9 (38.7) & 13.4 (6.6) \\
    & non-PD ($n$=14) & 70.5 (8.6) & 8 (57.2) & - & - \\
    \hline
    \multirow{2}{*}{\textit{Medell\'{i}n}} & PD ($n$=36) & 63.5 (8) & 12 (33) & 61.6 (37.4) & 30.8 (12) \\
    & non-PD ($n$=36) & 63.3 (6.2) & 12 (33) & - & - \\
    \hline
    \multirow{2}{*}{\textit{San Diego}} & PD ($n$=15) & 63.3 (8.2) & 8 (53.3) & 53.6 (40.5) & 32.7 (10.4) \\
    & non-PD ($n$=16) & 63.5 (9.7) & 9 (56.2) & - & - \\
    \hline
    \multirow{2}{*}{\textit{Turku}} & PD ($n$=19) & 69.6 (7.7) & 11 (57.9) & 80.5 (63) & 27.6 (16.9) \\
    & non-PD ($n$=19) & 67.5 (6.4) & 12 (63.2) & - & - \\
    \hline
\end{tabular}}
\vspace{-0.25cm}
\end{center}
\end{table}

We present each dataset's available demographic and clinical characteristics of PD and non-PD subjects in Table \ref{tab:datasets} as mean and standard deviation (SD). Worthy of note that there is no reported disease duration for two female subjects from the Turku dataset.

\subsubsection{Acquisition} 
\begin{figure}
    \centering
    \includegraphics[width=0.25\textwidth]{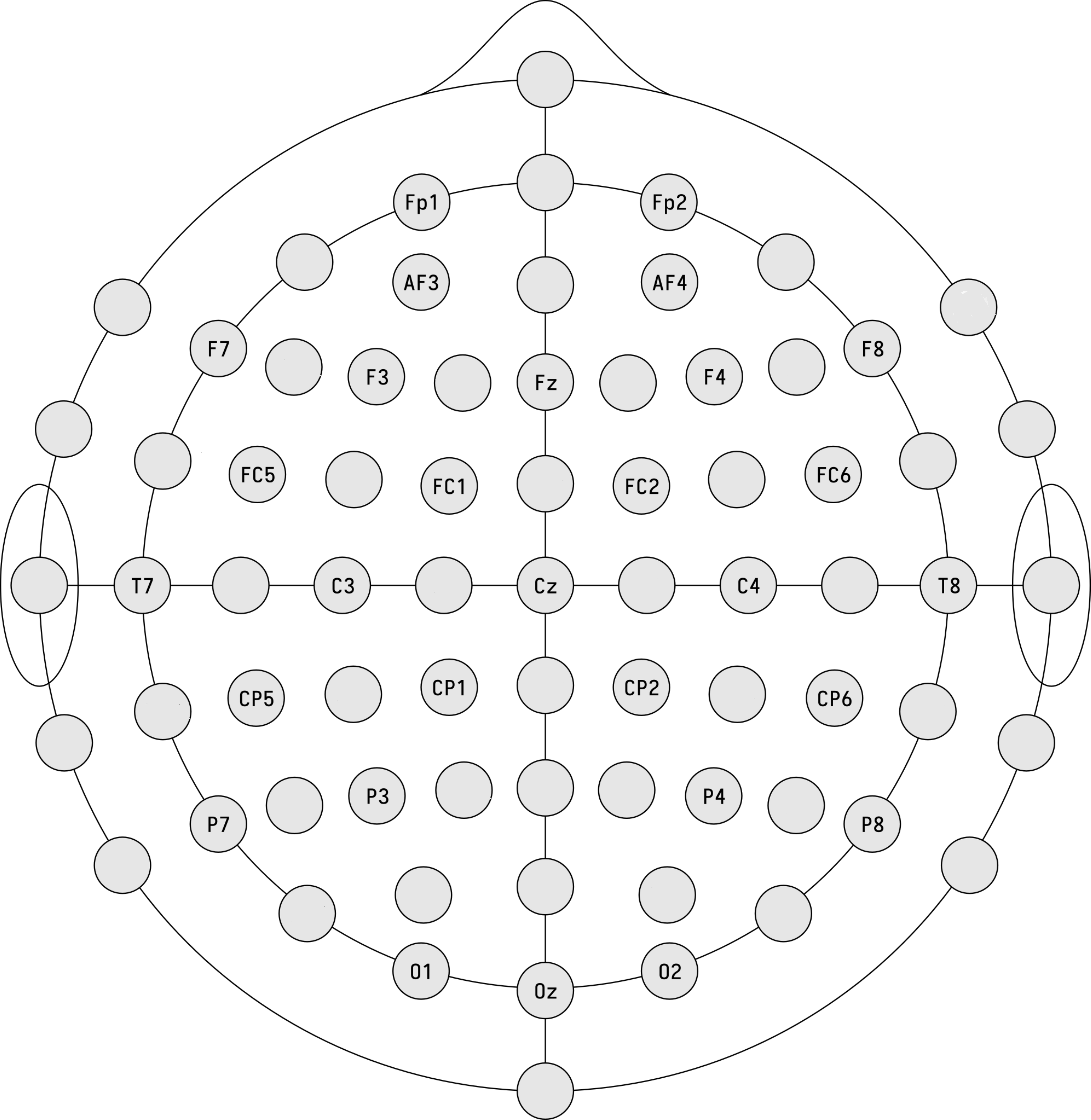}
    \caption{\small Common EEG channels included in the analysis.}
    \label{fig:channels}
    \vspace{-0.7cm}
\end{figure}

The rs-EEG recordings were acquired with the eyes closed for Medell\'{i}n, San Diego, and Turku subjects. In contrast, the subjects from Iowa City had their eyes open. For PD subjects from Iowa City, Medell\'{i}n, and San Diego, the rs-EEG recordings were from the ON phase of levodopa sessions, whilst, for Turku, only 6 PD subjects participated in EEG acquisition in the ON phase, while the remaining were in the OFF phase. All the datasets had 29 channels in common that were included in the analysis: AF3, AF4, C3, C4, CP1, CP2, CP5, CP6, Cz, F3, F4, F7, F8, FC1, FC2, FC5, FC6, Fp1, Fp2, Fz, O1, O2, Oz, P3, P4, P7, P8, T7, and T8 (Fig. \ref{fig:channels}).

\vspace{-0.1cm}
\subsubsection{Pre-processing} We applied a robust average re-referencing and adaptive line-noise correlation combined with detecting and interpolating noisy channels. The recordings were high-pass filtered at 1 Hz and decomposed using wavelet-based independent component (ICA) analysis to eliminate artifacts. We applied a low-pass filter at 30 Hz followed by segmentation of the signals into epochs of five-second length. Finally, we applied an additional rejection of artifact-compromised epochs. The common number of remaining epochs across all subjects within each dataset was included in the analysis.

\begin{table*}[!t]
\centering
\caption{\small Analysis of logistic regression (LR) model's performance (mean [95$\%$ CI]) for different gender sub-groups. Each result is presented in terms of global and center-wise performances, which are presented together with the number of subjects in their corresponding test sub-samples. f=females and m=males.}
\label{table:test}
\begin{tabular}{ c|c|c|c|c|c|c|c } 
\cline{2-8}
& & Accuracy, $\%$ & Recall, $\%$ & Specificity, $\%$ & Precision, $\%$ & $F_1$, $\%$ & AUC \\
\hline
\multirow{5}{*}{mixed} & global (19f, 32m) & \textbf{73.2 [66.5 79.9]} & \textbf{68.9 [59.8 78.0]} & \textbf{77.4 [67.3 87.5]} & \textbf{75.0 [66.2 83.8]} & \textbf{71.6 [64.4 78.7]} & \textbf{0.73 [0.66 0.80]} \\
& \textit{Iowa City} (3f, 5m) & 62.7 [52.2 73.3] & 25.5 [4.5 46.5] & 100 [100 100] & 78.0 [26.6 129.4] & 37.6 [9.4 65.8] & 0.63 [0.52 0.73] \\
& \textit{Medell\'{i}n} (5f, 17m) & 68.2 [57.2 79.2] & 72.6 [57.2 88.1] & 63.8 [46.5 81.2] & 67.5 [55.7 79.2] & 69.4 [58.5 80.3] & 0.68 [0.57 0.79] \\
& \textit{San Diego} (3f, 6m) & 77.3 [66.1 88.5] & 75.0 [75.0 75.0] & 79.2 [59.0 99.4] & 77.3 [57.8 96.9] & 75.4 [66.0 84.7] & 0.77 [0.67 0.87] \\ 
& \textit{Turku} (8f, 4m) & 84.2 [72.1 96.4] & 81.2 [62.5 99.8] & 87.3 [73.3 101] & 87.2 [73.6 101] & 83.3 [69.7 96.9] & 0.84 [0.72 0.96] \\
\hline
\multirow{5}{*}{males} & global (32m) & \textbf{80.5 [72.0 89.0]} & \textbf{88.4 [78.5 98.2]} & \textbf{74.3 [61.6 87.1]} & \textbf{73.6 [63.2 84.0]} & \textbf{80.0 [71.8 88.1]} & \textbf{0.81 [0.73 0.90]} \\
& \textit{Iowa City} (5m) & 80.6 [63.6 97.6] & 51.5 [9.0 94.0] & 100 [100 100] & 78.0 [26.6 130] & 60.3 [17.0 104] & 0.76 [0.54 0.97] \\
& \textit{Medell\'{i}n} (17m) & 70.9 [57.8 84.1] & 87.0 [71.9 102] & 59.7 [39.5 79.9] & 61.5 [48.4 74.5] & 71.4 [59.8 83.0] & 0.73 [0.61 0.86] \\
& \textit{San Diego} (6m) & 83.3 [66.8 99.9] & 100 [100 100] & 66.7 [33.6 99.8] & 78.1 [58.5 97.7] & 86.8 [74.6 99.1] & 0.83 [0.67 1.0] \\ 
& \textit{Turku} (4m) & 100 [100 100] & 100 [100 100] & 100 [100 100] & 100 [100 100] & 100 [100 100] & 1.0 [1.0 1.0] \\
\hline
\multirow{5}{*}{females} & global (19f) & \textbf{63.7 [54.1 73.3]} & \textbf{46.1 [32.0 60.2]} & \textbf{88.0 [74.8 101]} & \textbf{85.0 [68.9 101]} & \textbf{58.9 [45.6 72.3]} & \textbf{0.67 [0.58 0.76]} \\
& \textit{Iowa City} (3f) & 33.3 [33.3 33.3] & 0.0 [0.0 0.0] & 100 [100 100] & 0.0 [0.0 0.0] & 0.0 [0.0 0.0] & 0.50 [0.50 0.50] \\
& \textit{Medell\'{i}n} (5f) & 60.8 [38.1 83.5] & 51.0 [22.6 79.4] & 100 [100 100] & 94.0 [64.5 124] & 64.1 [35.4 92.8] & 0.76 [0.61 0.90] \\
& \textit{San Diego} (3f) & 66.7 [66.7 66.7] & 0.0 [0.0 0.0] & 100 [100 100] & 0.0 [0.0 0.0] & 0.0 [0.0 0.0] & 0.50 [0.50 0.50] \\ 
& \textit{Turku} (8f) & 74.5 [57.0 92.0] & 74.2 [48.1 100] & 74.8 [49.3 100] & 76.9 [55.5 98.2] & 73.7 [54.3 93.1] & 0.74 [0.57 0.92] \\
\hline

\end{tabular}
\vspace{-0.5cm}
\end{table*}

\vspace{-0.1cm}

\subsection{Parkinson's Disease (PD) Detection Model}

\vspace{-0.1cm}

A detailed description of the steps followed to obtain the PD detection model used in this work can be found in \cite{Kurbatskaya2023}. For the sake of simplicity, we provide a brief description to give a general idea of the model development and design:

\subsubsection{Feature Extraction and Harmonization}
For each epoch, relative PSD values were computed by dividing PSD values within a given frequency sub-band by the total signal power. In total, six sub-bands were applied: $\delta$ [1$-$4) Hz, $\theta$ [4$-$8) Hz (additionally divided into slow-$\theta$ [4$-$5.5) Hz and fast-$\theta$ [5.5$-$8) Hz, $\alpha$ [8$-$13) Hz, and $\beta$ [13$-$30) Hz. Besides, $\alpha$/$\theta$ PSD ratio is obtained. The resulting PSD values of all the epochs were further averaged, log-transformed, and harmonized using the modified ComBat harmonization model, producing 203 features (7 PSD values$\times$29 channels) per subject.  

\subsubsection{Classification Framework}
The overall sample was randomly divided into train and test sub-samples with a 70/30$\%$ split stratified by center and diagnosis (Fig. \ref{fig:distribution}). The best-performing model was selected from several ML classifiers based on the validation accuracy of a $subject$-wise binary classification of PD or non-PD obtained by a nested 5-fold cross-validation (CV) loop. In addition, we performed feature selection by means of a univariate $k$-feature selection based on ANOVA $F$-value. Model evaluation, hyperparameter space, and optimal number of features were evaluated in terms of average and 95$\%$ confidence intervals (CIs) for accuracy, recall, specificity, precision, $F_1$ score, and area under the receiver operating characteristic (ROC) curve (AUC) for the multi-center dataset. 


\vspace{-0.3cm}

\subsection{Fairness Analysis}

\subsubsection{Gender sub-group performance} We analyze the PD detection performance for male and female gender sub-groups (Fig. \ref{fig:distribution}). \textit{At this point, we would like to emphasize that developing a detection model is not the primary concern of the study, nor do we claim any contribution in that regard}. 

We start by replicating the original train and test sub-samples obtained by a 70/30$\%$ stratified split by center and diagnosis. We note that the original dataset has an equal presence of the gender sub-groups (Fig. \ref{fig:distribution}), resulting in a \textit{balanced representation of genders in both train and test sub-samples}. In the remainder of the article, we \textit{focus on the test sub-sample}. We perform the gender sub-group analysis by means of a \textit{stratified bootstrapping} approach with \textit{n=}100 replicates with replacement.

\textbf{Debiasing sub-group estimation via bootstrapping:} Limitations in some of the existing sub-group analyses include random splitting of the original dataset that was already presenting bias in the form of a higher prevalence of some of the sub-groups. We explore the use of gender, diagnosis, and center as stratification parameters to construct a test sub-group representative of the population of interest via bootstrapping. We perform \textit{n=}100 {replicates with replacement} and analyze the results for male and female sub-groups in terms of accuracy, recall, specificity, precision, $F_1$ score, and AUC. 

We present the results for each gender sub-group and the initial splitting results \cite{Kurbatskaya2023}. 
Further, we present the rounded confusion matrix for the original classifier, male sub-group, and female sub-group averaged over the \textit{n=}100 bootstrap replicates.
\begin{figure*}[!ht]
     \centering
     \begin{subfigure}[t]{0.3\textwidth}
         \centering
         \includegraphics[trim={0.7cm 0.2cm 2.3cm 1.2cm},clip,width=\textwidth]{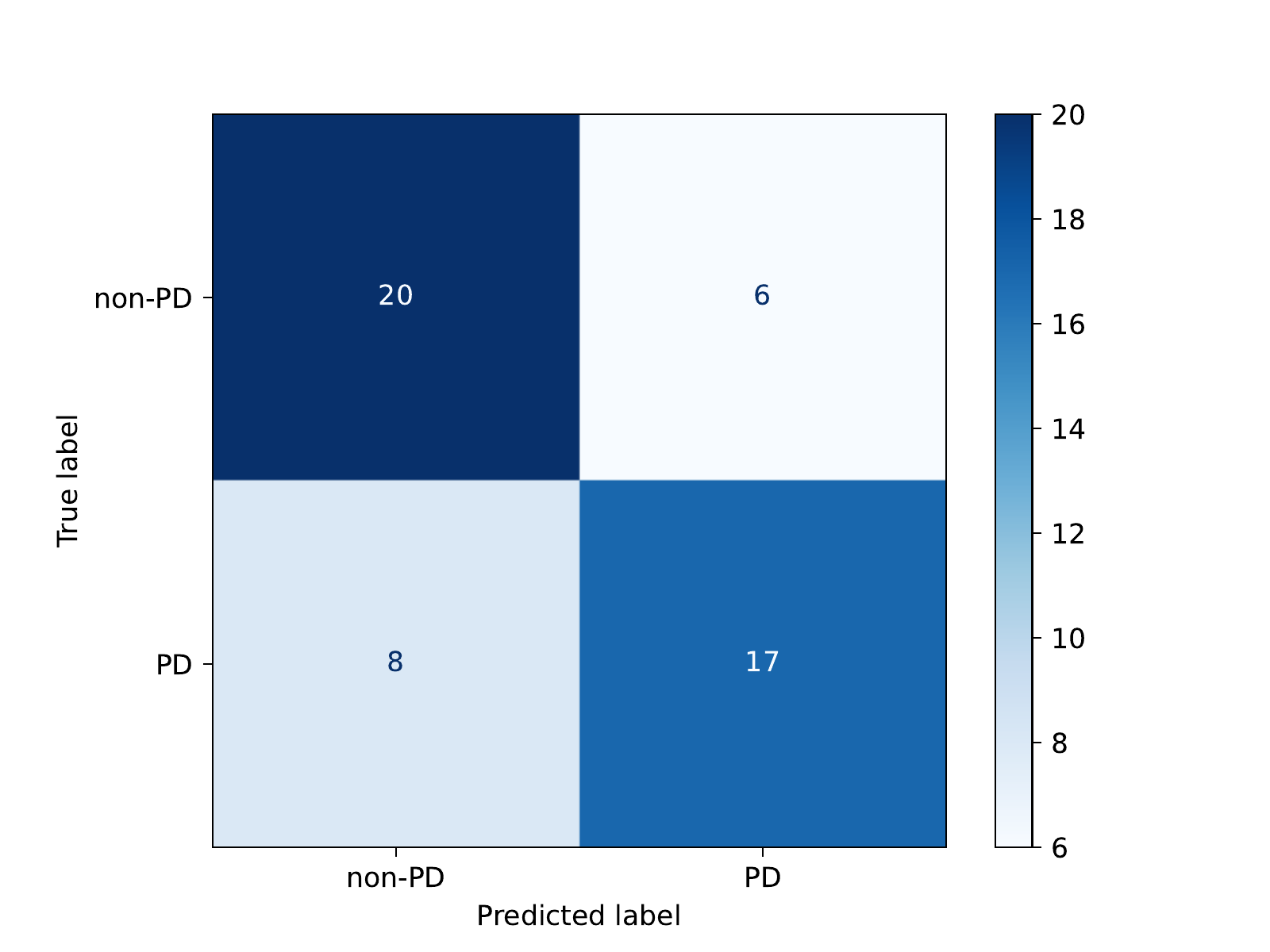}
         \caption{both genders}
         \label{fig:cm_both}
     \end{subfigure}
     \hfill
     \begin{subfigure}[t]{0.3\textwidth}
         \centering
         \includegraphics[trim={0.7cm 0.2cm 2.3cm 1.2cm},clip,width=\textwidth]{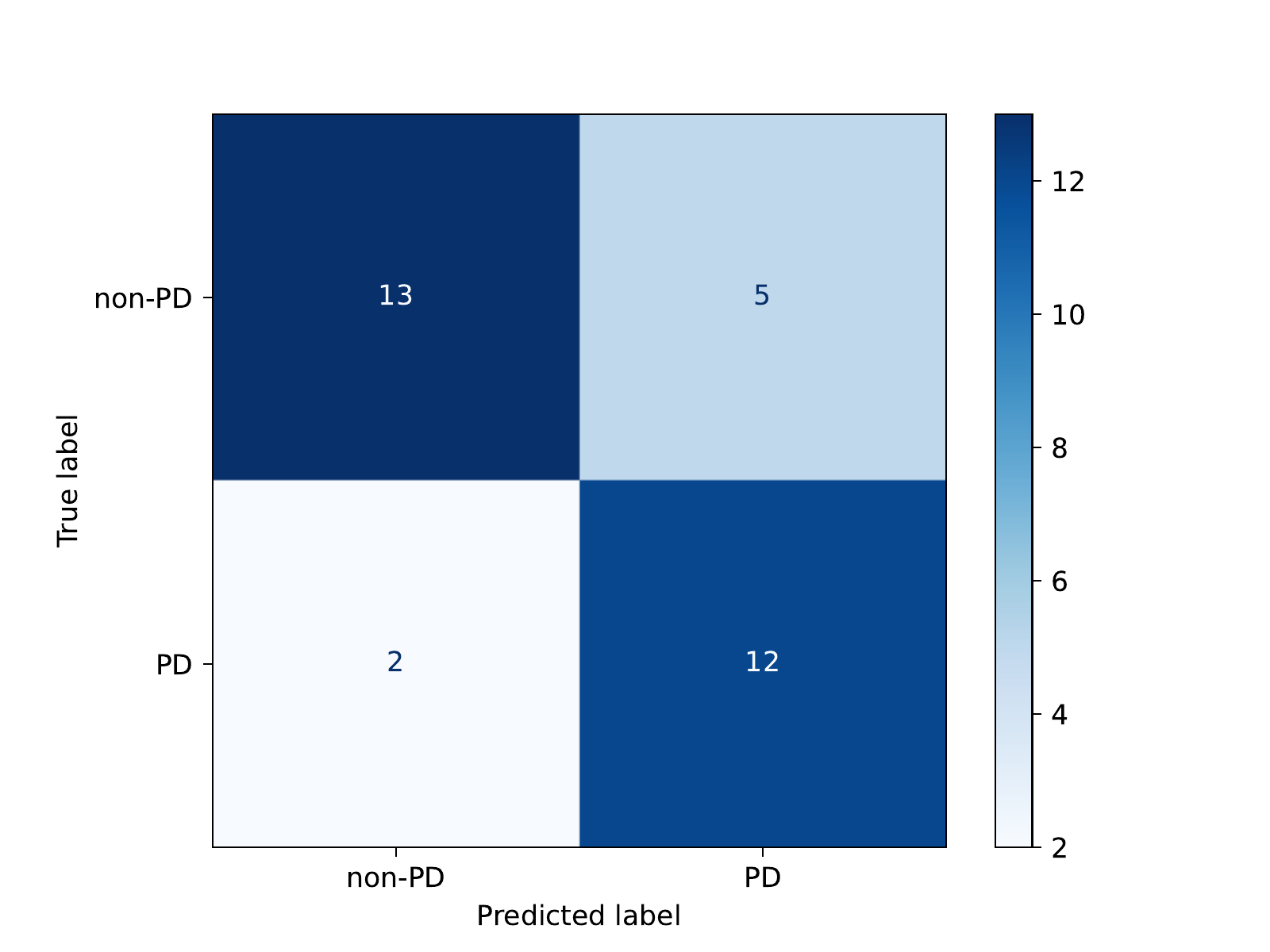}
         \caption{males}
         \label{fig:cm_males}
     \end{subfigure}
     \hfill
     \begin{subfigure}[t]{0.3\textwidth}
         \centering
         \includegraphics[trim={0.7cm 0.2cm 2.3cm 1.2cm},clip,width=\textwidth]{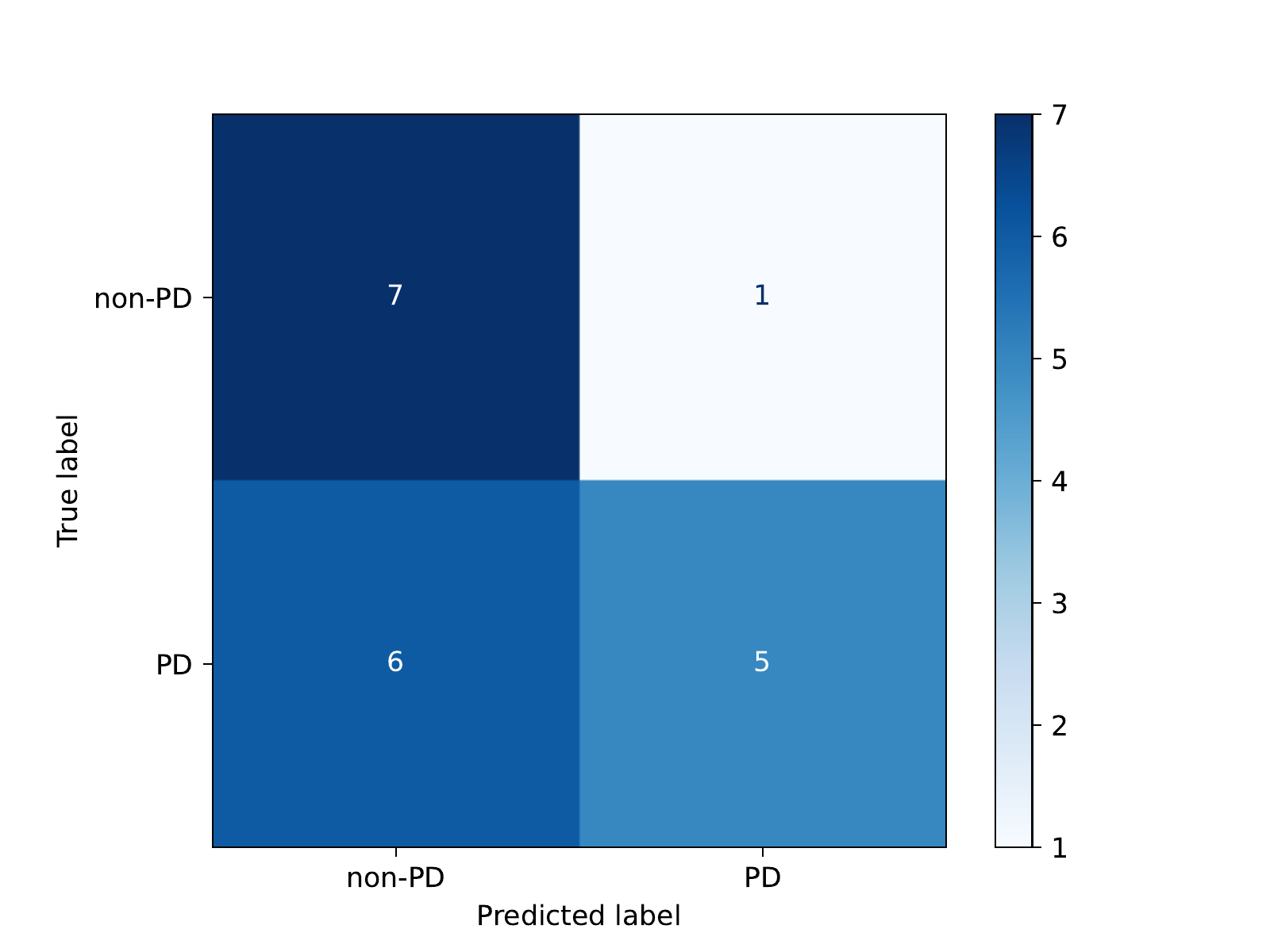}
         \caption{females}
         \label{fig:cm_females}
     \end{subfigure}
        \caption{\small Confusion matrices demonstrating the test accuracy of the trained logistic regression (LR) model to classify PD and non-PD subjects using mixed \ref{fig:cm_both}, only males \ref{fig:cm_males}, and only females \ref{fig:cm_females} test sub-samples.}
        \label{fig:cm}
\vspace{-0.5cm}
\end{figure*}

\begin{figure}[!hb]
\vspace{-0.5cm}
     \centering
     \begin{subfigure}[c]{0.27\textwidth}
         \centering
         \includegraphics[trim={0.2cm 0.2cm 0.1cm 0.2cm},clip,width=\textwidth]{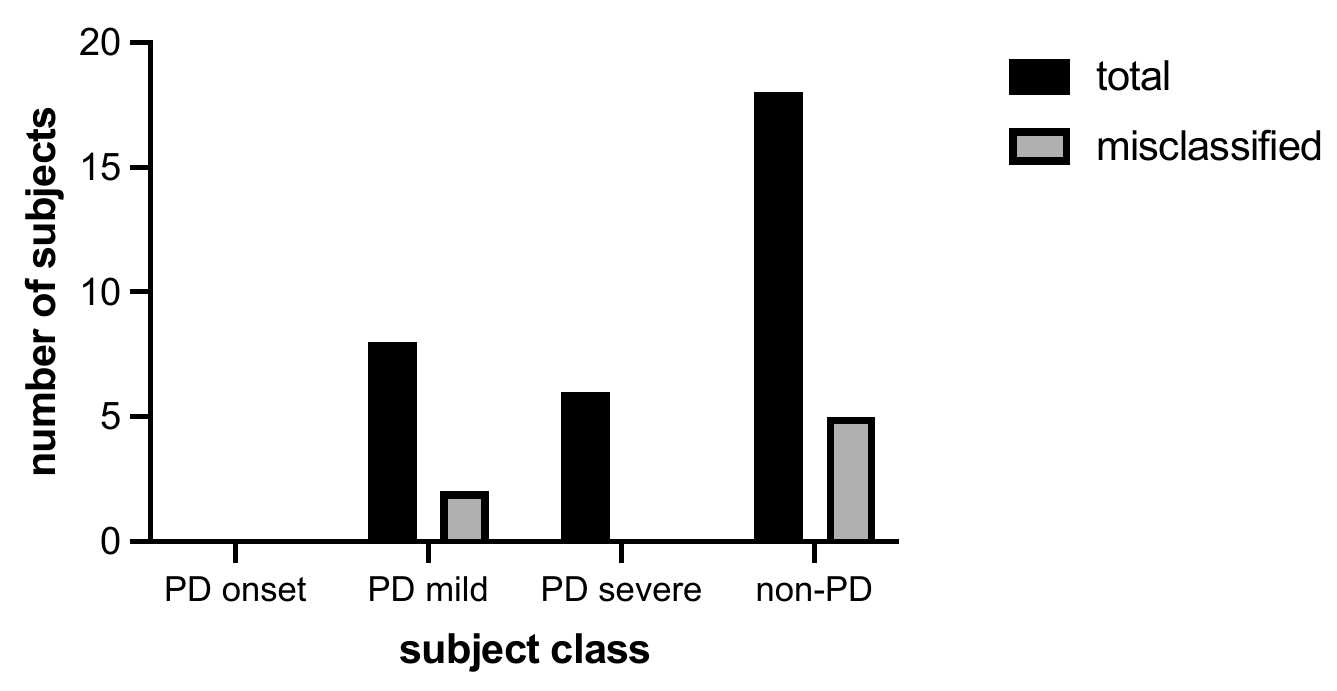}
         \label{fig:updrs_males}
     \end{subfigure}
     \hfill
     \begin{subfigure}[c]{0.20\textwidth}
         \centering
         \includegraphics[trim={0.2cm 0.2cm 4.0cm 0.2cm},clip,width=\textwidth]{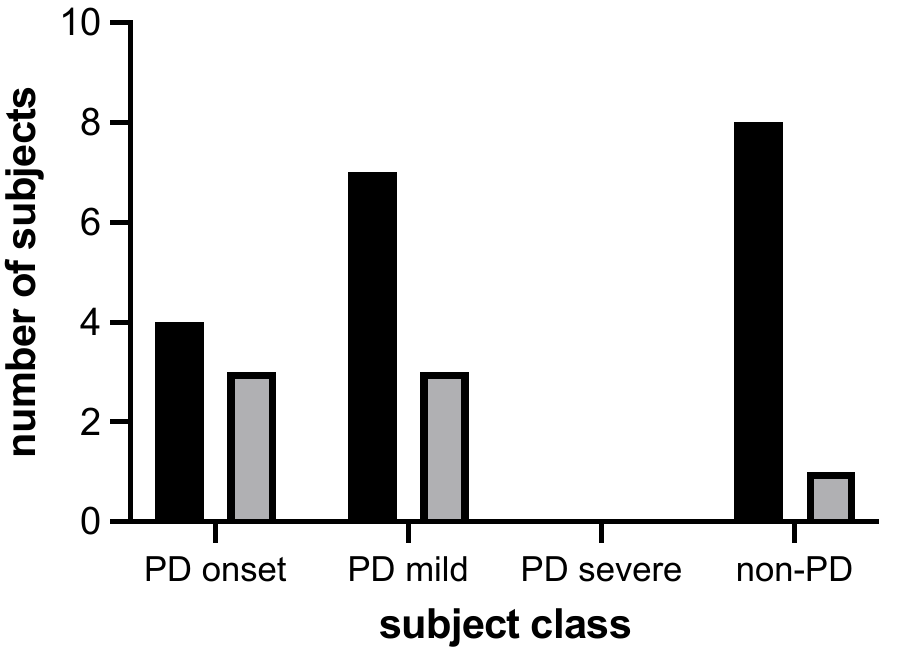}
         \label{fig:updrs_females}
     \end{subfigure}
    \hfill 
    \begin{subfigure}[c]{0.27\textwidth}
         \centering
         \includegraphics[trim={0.2cm 0.2cm 0.1cm 0.2cm},clip,width=\textwidth]{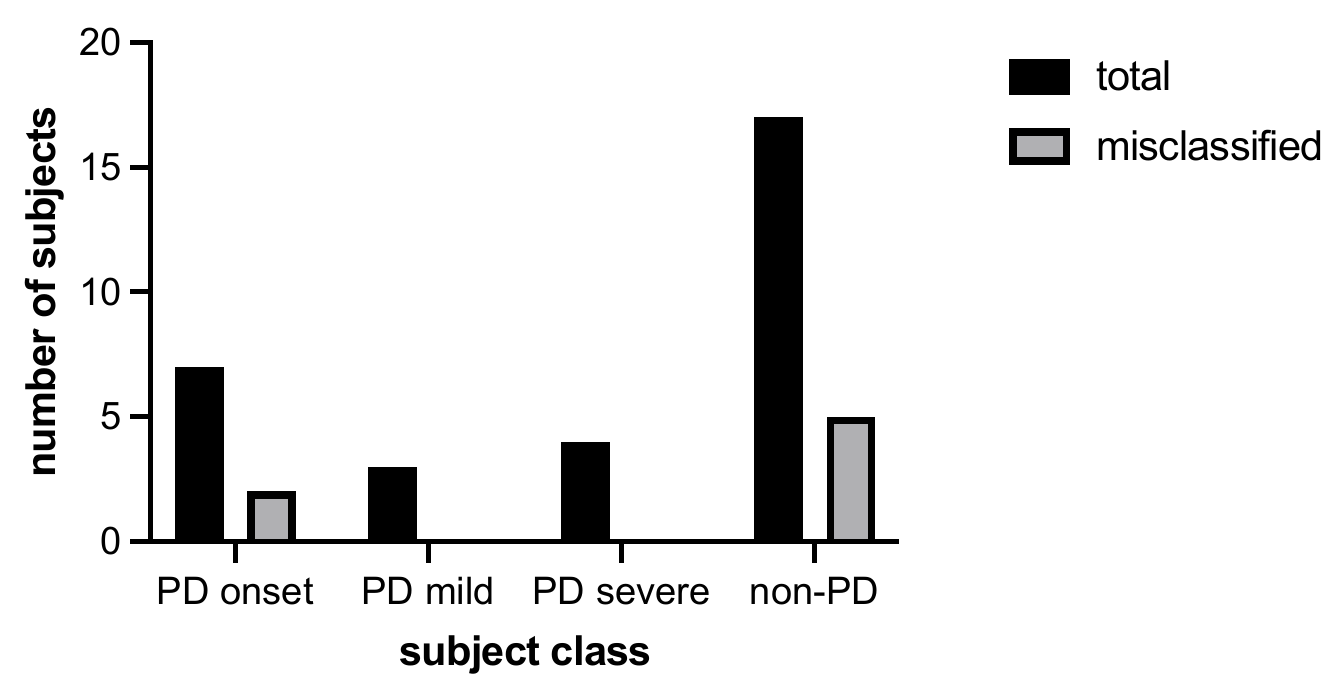}
         \caption{males}
         \label{fig:duration_males}
     \end{subfigure}
     \hfill
     \begin{subfigure}[c]{0.20\textwidth}  
         \centering
         \includegraphics[trim={0.2cm 0.2cm 4.0cm 0.2cm},clip,width=\textwidth]{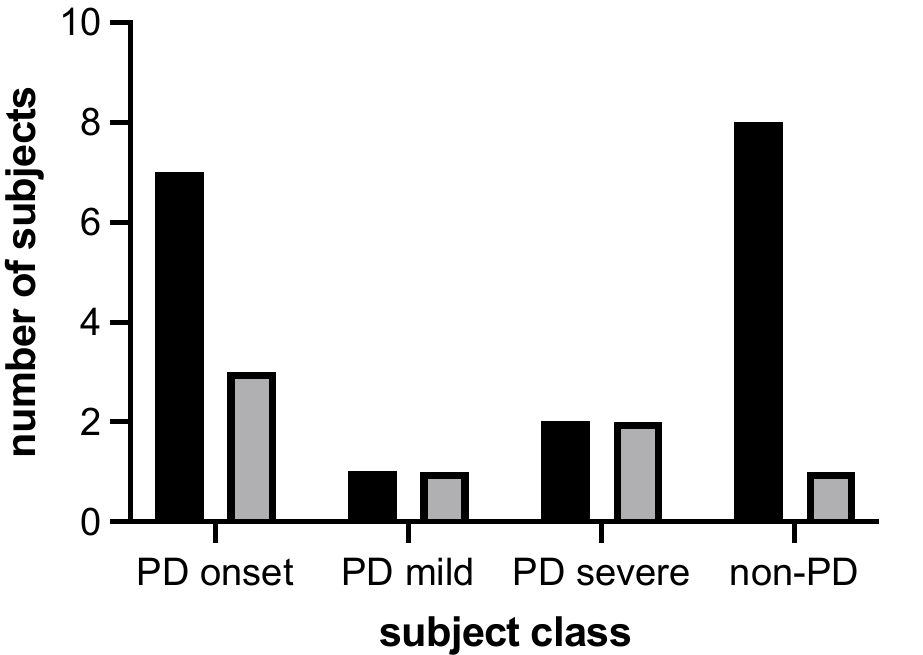}
         \caption{females}
         \label{fig:duration_females}
     \end{subfigure}
        \caption{\small Bar charts representing the number of misclassified PD divided into different groups based on UPDRS III scores (first row) and disease duration (second row). Misclassified non-PD subjects are also presented.}
        \label{fig:stages}
\vspace{-0.5cm}
\end{figure}

\subsubsection{Qualitative sub-group analysis of other attributes} We seek to understand whether misclassifications in gender sub-groups are correlated with differences in UPDRS III stage scores and disease duration. Since some of the included research centers use different versions of the UPDRS III stage, we  apply a simple conversion rule \cite{Hentz2015} to obtain Movement Disorder Society UPDRS (MDS-UPDRS) III from UPDRS III stage scores. Further, for the sake of simplicity, we trichotomize subjects based on the scores: $onset$ stage if score $\leqslant$ 20 \cite{Holden2018}, $mild$ stage if 20 $<$ score $<$ 35, and $severe$ stage if score $\geqslant$ 35. Similarly, subjects are trichotomized based on the PD duration (in months): $onset$ stage if duration $\leqslant$ 50 months, $mild$ stage if 50 $<$ duration $<$ 100 months, and $severe$ stage if duration $\geqslant$ 100 months. 

\subsubsection{rs-EEG feature relevance for gender sub-groups} In the last step of our fairness analysis, we focus on explaining the gender sub-group differences through the expressiveness in the detection framework of the PSD features extracted from the rs-EEG for each sub-group. With that in mind, we \textit{re-train the original model with a data sub-set containing only males, and we repeat the same process for females, obtaining two models specialized in each gender sub-group}. The proportions of the training and testing datasets are kept at 70/30$\%$, and we apply the original model development stratification strategy: center and diagnosis. We fix the male and female model hyperparameters and \textit{focus, exclusively, on the feature selection process} by means of ANOVA \textit{F}-value. We obtain the $k$-highest scoring features for the gender-specialized models with the best validation results through a 5-fold CV process. 

At this point, given our original biased model and our gender-specific models that keep all the original model design choices except for the feature selection process, \textit{we hypothesize that there are significant differences in rs-EEG features from the frontal and parietal regions of the brain \cite{Bin2018, Teramoto2016} and that the three classifiers commonly select these features as high-scoring features for the PD classification}. Further, we also \textit{hypothesize that the features that present differences in average values between male PD, female PD, male non-PD, and female non-PD can explain the differences in sub-group performance in the original classifier}.

To corroborate our hypotheses, we perform $t$-test followed by a false discovery rate (FDR) correction for male and female sub-groups of the common $k$-highest scoring features from the frontal and parietal 
regions of the brain. Before accounting for corrections, we consider \textit{p} $<$ 0.05 to be significant. We apply FDR at the 0.05 significance level. Finally, we analyze the rs-EEG features with respect to the scalp position (channel position) and PSD sub-bands. We present the results in terms of mean and 95$\%$ CI. 

\vspace{-0.2cm}

\section{Results and Discussion}

As depicted in Table \ref{table:test}, we present the bootstrapped sub-group PD detection analysis in terms of the original classifier with matched genders and gender as a \textit{protected attribute}, only male and only female sub-group subjects. We present the results for all the datasets together (global) and the subjects of each dataset separately. Overall, there is an evident degradation in PD detection performance for females compared to males, reflected by a lower accuracy, recall, $F_1$ score, and AUC. The same trend is observed for all the individual contributing centers. Confusion matrices (Fig. \ref{fig:cm}) reveal that the model's PD detection for the male sub-group results in a high type I error (28$\%$), two times as high as the type I error in the female sub-sample (13$\%$). In comparison, detecting PD in the female sub-group results in a high type II error (55$\%$), almost four times as high as the type II error in the male sub-group (15$\%$).

The qualitative analysis presented in Fig. \ref{fig:stages} reveals that for males, the misclassified PD subjects belong to either the mild stage in terms of disease duration or onset stage based on UPDRS III scores. All the subjects from the severe stage groups are correctly detected by the model. This might indicate a lower model's detection of early-mild stages of PD for males. On the other hand, the misclassified PD female subjects have either the onset or mild stages based on UPDRS III scores and onset, mild or severe stages based on the disease duration. Regardless of the disease stage, the model presents a clear underperformance for PD detection in female subjects when compared to males.

The best validation results are achieved with \textit{k=}188 and \textit{k=}181 highest scoring features for the male and female gender sub-group models, respectively. Out of these features, 169 are selected as common features for both models, and channels from both the frontal and parietal brain regions are further analyzed to limit pairwise comparisons and on the basis of previous research results. Table \ref{table:features} presents mean and 95$\%$ CI values of the \textit{significantly different}  (\textit{before FDR correction}) common PSD features with their corresponding channel and frequency sub-band for PD and non-PD samples for each gender sub-group. We note that after FDR correction, no significance is found in the differences. 

Given the log-transformation of our values and the amplitude range of the original rs-EEG recordings, higher mean absolute values might be indicative of lower activity for PD and non-PD in the channels and sub-bands whilst smaller mean absolute values might be indicative of higher activation in the channel and sub-band under analysis. In particular, we can observe that overall, PD males have \textit{higher activity} when compared to PD females in the parietal and frontal channels and resulting sub-bands of the analysis (Table \ref{table:features}). In a similar way, non-PD males show a \textit{higher activity} in selected channels and sub-bands when compared to non-PD females (Table \ref{table:features}), but not by a large margin.

\begin{table}[!hb]

\caption{\small Mean$\pm$SD values of the significantly different common features with their respective channels and frequency sub-bands for male and female subjects. $p$-values are presented non-corrected.}
\label{table:features}
\begin{center}
\notsotiny
\begin{tabular}{ c|c|c|c|c|c } 
\cline{2-6}
& Channel & Sub-band & Gender & Mean$\pm$SD & 95$\%$ CI \\
\hline
\multirow{34}{*}{PD} & \multirow{8}{*}{CP1} & \multirow{2}{*}{$\theta$} & male & -0.77$\pm$0.13 & [-0.93 -0.61] \\ 
& & & female & -0.96$\pm$0.22 & [-1.24 -0.69] \\  
\cline{3-6}
 & & \multirow{2}{*}{fast-$\theta$} & male & -0.91$\pm$0.17 & [-1.12 -0.71] \\
& & & female & -1.13$\pm$0.23 & [-1.41 -0.85] \\  
\cline{3-6}
 & & \multirow{2}{*}{$\beta$} & male & -0.67$\pm$0.25 & [-0.97 -0.36] \\
& & & female & -0.45$\pm$0.19 & [-0.68 -0.22] \\ 
\cline{3-6}
 & & \multirow{2}{*}{$\alpha/\theta$} & male & 0.26$\pm$0.25 & [-0.06 0.56] \\
& & & female & 0.47$\pm$0.24 & [0.17 0.76] \\ 

\cline{2-6}
 & \multirow{8}{*}{CP2} & \multirow{2}{*}{$\theta$} & male & -0.88$\pm$0.27 & [-1.22 -0.55] \\
& & & female & -1.00$\pm$0.25 & [-1.31 -0.70] \\ 
\cline{3-6}
 & & \multirow{2}{*}{fast-$\theta$} & male & -1.00$\pm$0.31 & [-1.38 -0.62] \\
& & & female & -1.14$\pm$0.27 & [-1.47 -0.81] \\
\cline{3-6}
 & & \multirow{2}{*}{$\beta$} & male & -0.62$\pm$0.30 & [-0.99 -0.24] \\
& & & female & -0.42$\pm$0.15 & [-0.60 -0.23] \\ 
\cline{3-6}
 & & \multirow{2}{*}{$\alpha/\theta$} & male & 0.27$\pm$0.26 & [-0.05 0.59] \\
& & & female & 0.49$\pm$0.22 & [0.22 0.75] \\

\cline{2-6}
 & \multirow{2}{*}{CP5} & \multirow{2}{*}{$\alpha/\theta$} & male & -0.72$\pm$0.16 & [-0.91 -0.53] \\
& & & female & -0.89$\pm$0.30 & [-1.27 -0.52] \\

\cline{2-6}
 & \multirow{2}{*}{CP6} & \multirow{2}{*}{$\alpha/\theta$} & male & -0.90$\pm$0.22 & [-1.17 -0.64] \\
& & & female & -1.00$\pm$0.26 & [-1.35 -0.71] \\ 

\cline{2-6}
 & \multirow{2}{*}{Fp1} & \multirow{2}{*}{$\alpha/\theta$} & male & -0.81$\pm$0.25 & [-1.11 -0.50] \\
& & & female & -0.62$\pm$0.24 & [-0.93 -0.32] \\ 

\cline{2-6}
 & \multirow{4}{*}{Fp2} & \multirow{2}{*}{$\alpha$} & male & -0.84$\pm$0.16 & [-1.05 -0.63] \\
& & & female & -0.99$\pm$0.29 & [-1.35 -0.63] \\ 
\cline{3-6}
 & & \multirow{2}{*}{$\alpha/\theta$} & male & 0.21$\pm$0.32 & [-0.19 0.61] \\
& & & female & 0.43$\pm$0.37 & [0.03 0.89] \\ 

\cline{2-6}
 & \multirow{8}{*}{P3} & \multirow{2}{*}{$\theta$} & male & -0.56$\pm$0.29 & [-0.92 -0.21] \\
& & & female & -0.46$\pm$0.16 & [-0.66 -0.26] \\ 
\cline{3-6}
 & & \multirow{2}{*}{slow-$\theta$} & male & 0.30$\pm$0.30 & [-0.07 0.66] \\
& & & female & 0.45$\pm$0.23 & [0.17 0.73] \\ 
\cline{3-6}
& & \multirow{2}{*}{fast-$\theta$} & male & -0.79$\pm$0.16 & [-0.99 -0.59] \\
& & & female & -0.92$\pm$0.20 & [-1.17 -0.67] \\ 
\cline{3-6}
 & & \multirow{2}{*}{$\beta$} & male & -0.91$\pm$0.18 & [-1.13 -0.69] \\
& & & female & -1.04$\pm$0.24 & [-1.33 -0.75] \\ 
\hline

\multirow{6}{*}{non-PD} & \multirow{2}{*}{CP5} & \multirow{2}{*}{$\delta$} & male & -1.11$\pm$0.30 & [-1.48 -0.74] \\ 
& & & female & -1.09$\pm$0.15 & [-1.28 -0.90] \\  
\cline{2-6}
\cline{2-6}
 & \multirow{2}{*}{Fp1} & \multirow{2}{*}{$\beta$} & male & -1.10$\pm$0.26 & [-1.42 -0.78] \\
& & & female & -1.12$\pm$0.17 & [-1.33 -0.92] \\ 
\cline{2-6}
 & \multirow{2}{*}{Fp2} & \multirow{2}{*}{slow-$\theta$} & male & -0.97$\pm$0.23 & [-1.26 -0.69] \\
& & & female & -0.98$\pm$0.12 & [-1.12 -0.84] \\ 
\hline
\end{tabular}
\vspace{-0.6cm}
\end{center}

\end{table}

\section{Conclusion}

Different motor and non-motor profiles of PD and non-PD in men and women might induce different rs-EEG profiles, which are harder to detect in women due to their overall lower activity in parietal and frontal channels. Our findings corroborate the exacerbation of gender disparities found in other studies. As depicted by our systematic analysis, measures to ensure fair ML development for PD detection are required. The findings can help to provide a more personalized diagnosis using rs-EEG and ML by accounting for gender differences in rs-EEG parietal and frontal activity. Limitations of the study include the use of retrospective data, a limited amount of patients, and statistical power.

\footnotesize {
\section*{Acknowledgments}

We would like to thank all the researchers and staff at the University of California San Diego, the University of Turku and the Turku University Hospital, the University of Iowa, and the University of Antioquia. We especially thank Dr. A. P. Rockhill, Dr. H. Railo, and Dr. N. S. Narayanan for making their datasets publicly available.}

\printbibliography[title=References]

\end{document}